\documentclass[aps,pra,twocolumn,floatfix]{revtex4}
\usepackage{graphicx}
\begin{document}
\title{Dramatic reductions in inelastic cross sections
\\ for ultracold collisions near Feshbach resonances}
\author{Jeremy M. Hutson}
\affiliation{Department of Chemistry, University of Durham,
South Road, Durham, DH1~3LE, England}

\author{Musie Beyene}
\affiliation{Department of Chemistry, University of Durham,
South Road, Durham, DH1~3LE, England}
\author{Maykel Leonardo
Gonz{\'a}lez-Mart{\'\i}nez} \affiliation{Departamento de
F{\'\i}sica General, InSTEC, Habana 6163, Cuba}

\date{\today}

\begin{abstract}
We show that low-energy inelastic cross sections can decrease
as well as increase in the vicinity of a zero-energy Feshbach
resonance. When an external field is used to tune across such a
resonance, the real and imaginary parts of the scattering
length show asymmetric oscillations, with both peaks and
troughs. In favorable circumstances the inelastic collision
rate can be reduced almost to zero. This may be important for
efforts to achieve evaporative and sympathetic cooling for
molecules.
\end{abstract}


\maketitle



Cold and ultracold molecules have fascinating properties that
will find applications in many areas of physics, ranging from
precision measurement to quantum computing
\cite{Carr:NJPintro:2009}. Ultracold molecules offer new
possibilities for quantum simulations and quantum control,
while quantum gases of ultracold polar molecules are expected
to exhibit a wide range of new quantum phases.

Cold and ultracold molecules must always be confined in traps,
and trap losses are crucial. In particular, collisional
stability is very important. Magnetic and electrostatic traps
can trap molecules only when they are in low-field-seeking
states, and such states are never the lowest in the applied
field. Any inelastic collision that transfers internal energy
into relative translational motion causes either heating or
trap loss. It is thus very important to understand inelastic
collisions and to find ways to minimise them. The purpose of
the present paper is to show that inelastic collision rates can
sometimes be dramatically reduced by tuning close to a Feshbach
resonance \cite{Feshbach:1962} with an applied electric or
magnetic field.

It has been possible for about 5 years to create molecules in
highly vibrationally excited states in ultracold atomic gases
\cite{Hutson:IRPC:2006}, both by photoassociation
\cite{Jones:RMP:2006} and by magnetoassociation
\cite{Kohler:RMP:2006}. A major goal was achieved in 2008 when
Ni {\em et al.}\ \cite{Ni:KRb:2008} succeeded in transfering
KRb molecules formed by magnetoassociation at $T=350$~nK into
their ground rovibrational state by stimulated Raman adiabatic
passage (STIRAP). Experiments on collisional trap loss are
already under way \cite{Ye:private:2009}. Danzl {\em et al.}\
\cite{Danzl:v73:2008, Danzl:ground:2009} and Lang {\em et al.}\
\cite{Lang:ground:2008} have carried out analogous experiments
on Cs$_2$ and triplet Rb$_2$, respectively. There have also
been considerable successes in direct photoassociation to
produce low-lying states \cite{Sage:2005, Hudson:PRL:2008,
Viteau:2008, Deiglmayr:2008}.

Methods based on photoassociation and magnetoassociation are
limited to molecules formed from atoms that can be
laser-cooled, such as the alkali metals. However, a wider range
of molecules can be cooled directly from high temperature to
the millikelvin regime, using methods such as buffer-gas
cooling \cite{Weinstein:CaH:1998} and Stark deceleration
\cite{Bethlem:IRPC:2003}. Polar molecules such as ND$_3$ and OH
can be held in electrostatic or alternating current traps
\cite{Bethlem:trap:2000, vanVeldhoven:2005}, while paramagnetic
molecules such as CaH, O$_2$, NH and OH can be held in magnetic
traps. However, at present the lowest temperatures that can be
achieved in static traps are around 10 mK, and there are a
variety of proposals for ways to cool them further, including
evaporative cooling, sympathetic cooling and cavity-assisted
cooling \cite{Domokos:2002, Chan:2003, Morigi:2007}.

Magnetic fields have important effects on the interactions and
collisions of paramagnetic molecules \cite{Volpi:2002,
Krems:henh:2003, Krems:IRPC:2005}. In previous work
\cite{Gonzalez-Martinez:2007}, we explored the use of Feshbach
resonances to {\it control} molecular collisions with applied
fields. For the prototype system He + NH ($^3\Sigma^-)$, we
located magnetic fields at which bound states cross
open-channel thresholds and then characterized the resulting
low-energy Feshbach resonances as a function of magnetic field.
For a resonance at which a bound state crosses the {\it lowest}
open-channel threshold, the real scattering length $a$ behaves
in the same way as in the atomic case \cite{Moerdijk:1995} and
exhibits a pole as a function of magnetic field $B$. However,
for resonances in which a state crosses a {\it higher}
threshold, we observed quite different behavior. In this case,
inelastic scattering can occur and the scattering length is
complex. The complex scattering length $a(B)=\alpha(B)-{\rm
i}\beta(B)$ was found to follow the formula
\cite{Gonzalez-Martinez:2007}
\begin{equation}
a(B) = a_{\rm bg} + \frac{a_{\rm res}}{2(B-B_{\rm
res})/\Gamma_B^{\rm inel}+{\rm i}}, \label{eqaares}
\end{equation}
where $a_{\rm bg}$ is a slowly varying background term and
$B_{\rm res}$ and $\Gamma_B^{\rm inel}$ are the position and
width of the resonance. The {\it resonant scattering length}
$a_{\rm res}$ characterizes the strength of the resonance. The
elastic and total inelastic cross sections are given
approximately by
\begin{equation}
\sigma_{\rm el}(B) \approx {4\pi|a(B)|^2} \hbox{\quad
and \quad} \sigma_{\rm inel}^{\rm tot}(B) \approx \frac{4\pi\beta(B)}{k_0}.
\end{equation}

The inelastic scattering in He + NH is very weak except near
resonance. Under these circumstances, $\sigma_{\rm el}(B)$
shows a symmetric oscillation at resonance and $\sigma_{\rm
inel}^{\rm tot}(B)$ shows a simple peak
\cite{Gonzalez-Martinez:2007}. This corresponds to a real value
of $a_{\rm res}$. However, a more complete derivation
\cite{Hutson:res:2007} subsequently showed that, when the
background scattering is significantly inelastic, it is
possible for $a_{\rm res}$ to be complex. When this occurs,
inelastic cross sections show troughs as well as peaks near
resonance. This is potentially of great importance, since
inelastic collisions generally provide trap loss mechanisms and
strong inelastic processes can prevent evaporative or
sympathetic cooling.

In this letter we consider a more strongly coupled system, with
significant inelastic scattering, in order to demonstrate the
dramatic reductions in inelasticity that can occur near
Feshbach resonances. The system we have chosen is $^4$He +
$^{16}$O$_2$ ($^3\Sigma_g^-$), for which a reliable potential
energy surface has been calculated by Groenenboom and
Struniewicz \cite{Groenenboom:2000}. The $^{16}$O$_2$ molecule
has a ground state with rotational quantum number $n=1$,
because the $^{16}$O nucleus is a boson with nuclear spin
$I=0$. The bound-state Schr\"odinger equation for $^4$He +
$^{16}$O$_2$ is solved by propagating coupled differential
equations using the BOUND package
\cite{Hutson:bound-short:1993, Hutson:cpc:1994}, as modified to
handle magnetic fields \cite{Gonzalez-Martinez:2007}. The
calculations are carried out in a completely decoupled basis
set \cite{Krems:henh:2003}, $|nm_n\rangle |sm_s\rangle
|Lm_L\rangle$, where $s=1$ is the electron spin of O$_2$ and
$L$ is the end-over-end rotational angular momentum of He and
the molecule. All the $m$ quantum numbers represent space-fixed
projections on the axis defined by the magnetic field. The only
good quantum numbers are the parity $(-1)^{n+L+1}$ and the
total projection quantum number $M_{\rm tot} = m_n+m_s+m_L$.
At energies above the lowest threshold, BOUND locates both
physical quasibound states and artificial states that result
from box-quantizing the continuum for open channels. However,
it is straightforward to identify the physical states by
inspecting the dependence of the eigenvalues on the outer limit
of the propagation.

\begin{figure}
\includegraphics[width=0.98\linewidth]{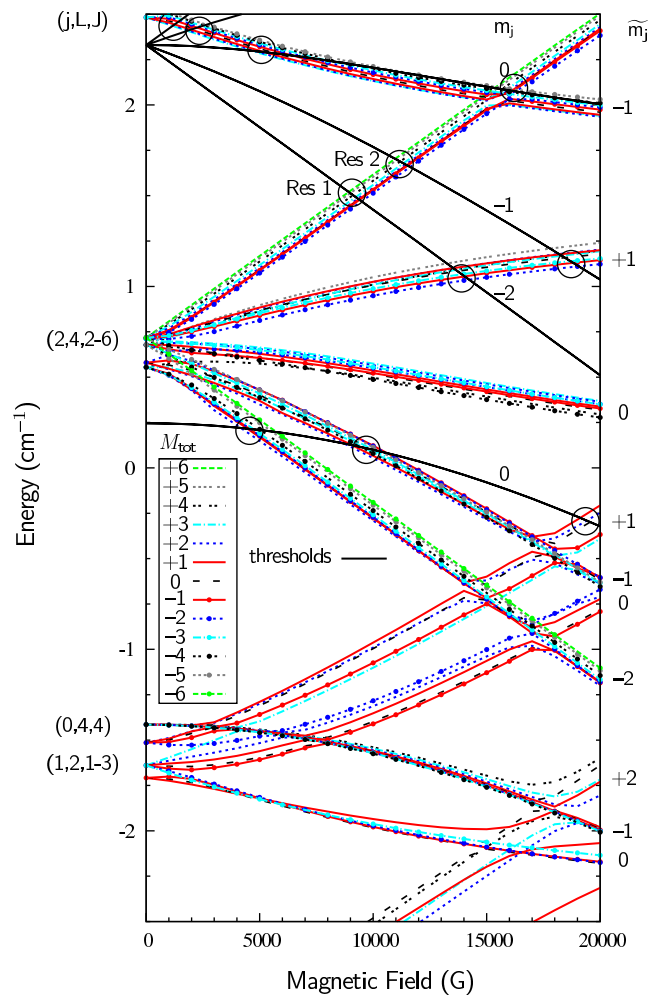}
\caption{(Color online). The pattern of levels from bound-state
calculations on $^4$He-$^{16}$O$_2$ near the $n=1$ thresholds,
with artificial levels removed, as a function of magnetic field $B$.
The calculations are for even parity, $M_{\rm tot}=-6$ to $+6$.
The $^{16}$O$_2$ threshold energies are shown as solid black lines.
The circles show crossings between bound states and thresholds
with $m_j=M_{\rm tot}$ that produce zero-energy Feshbach
resonances in s-wave scattering.
}
\label{fig-levs-B}
\end{figure}

Fig.\ \ref{fig-levs-B} shows the bound and quasibound states of
the $^4$He-$^{16}$O$_2$ complex near the $n=1$ thresholds as a
function of magnetic field $B$, with artificial levels removed,
together with the thresholds for dissociation to form He +
O$_2$. The thresholds are characterized at zero field by O$_2$
quantum numbers $n,s,j$ with $s=1$ and $j=0,1,2$ for $n=1$, and
each one splits into $2j+1$ components labeled by $m_j$ at
non-zero magnetic field. He-O$_2$ is a weakly anisotropic
system, so $n$ and $s$ remain nearly good quantum numbers and
the levels of the complex are characterized by additional
quantum numbers $L$ and $J$, where the total angular momentum
$J$ is the resultant of $j$ and $L$. At zero field each
$(n,j,L)$ level splits into $\min(2j+1,2L+1)$ sublevels with
different values of $J$. When a magnetic field is applied, each
sublevel splits into $2J+1$ components with different values of
$M_{\rm tot}$. The $J$ quantum number remains a useful label
for magnetic fields up to about 1000 G, but above that the
levels of different $J$ are strongly mixed. By about 5000 G the
levels have separated into groups that may be labeled with an
approximate quantum number $\widetilde m_j$ that takes values
from $-j$ to $+j$.

Crossings between quasibound states and thresholds will produce
zero-energy Feshbach resonances in s-wave scattering if an
$L=0$ scattering channel is permitted by the constraints on
parity and $M_{\rm tot}$. This occurs only for thresholds
corresponding to $m_j=M_{\rm tot}$ as shown by the circles in
Fig.\ \ref{fig-levs-B}. In the present work we are particularly
interested in resonances that occur at {\em excited}
thresholds, where inelastic scattering may occur.

Once the crossing points have been located in Fig.\
\ref{fig-levs-B}, we carry out scattering calculations, holding
the kinetic energy fixed at a small value while sweeping the
magnetic field across the resonance. This is done using the
MOLSCAT package \cite{molscat:v14-short}, as modified to handle
collisions in magnetic fields \cite{Gonzalez-Martinez:2007}.

\begin{figure}[tb]
\begin{center}
\includegraphics[width=0.95\linewidth]{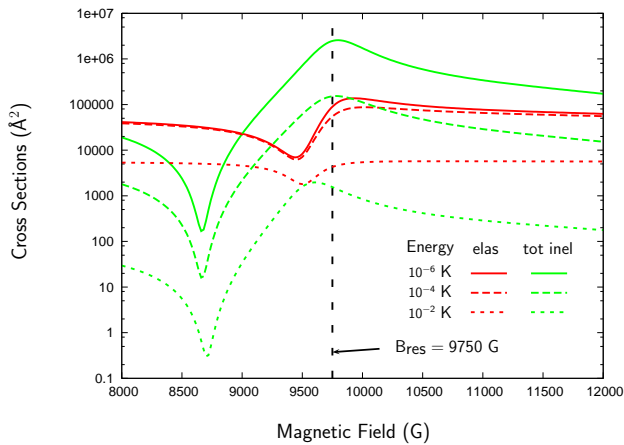}
\caption{(Color online). Elastic (red) and total inelastic
(green) cross sections for resonance 1 in He + O$_2$ as
a function of magnetic field at collision energies of $10^{-6}$~K
(solid lines), $10^{-4}$~K (dashed lines) and $10^{-2}$~K
(dotted lines).}
\label{fig-reso2-1}
\end{center}
\end{figure}

\begin{figure}[tb]
\begin{center}
\includegraphics[width=0.95\linewidth]{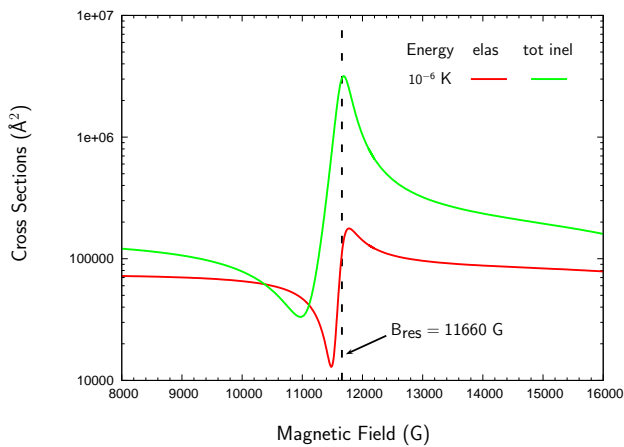}
\caption{(Color online). Elastic (red) and total inelastic
(green) cross sections for resonance 2 in He + O$_2$ as
a function of magnetic field at collision energy $10^{-6}$~K.}
\label{fig-reso2-2}
\end{center}
\end{figure}

Typical resonance profiles for $^4$He + O$_2$ are shown in
Fig.\ \ref{fig-reso2-1}, for the resonance labeled 1 in Fig.\
\ref{fig-levs-B} at collision energies $E=1$~$\mu$K, 100~$\mu$K
and 10~mK. At all three energies the total inelastic cross
section (summed over outgoing partial waves $L'$) drops by
almost a factor of 1000 from its background value at a field
just below $B_{\rm res}$. At higher energies the resonance is
shifted slightly, but the resonant suppression is just as
strong. The resonant contribution to the inelastic cross
section follows the $E^{-1/2}$ Wigner threshold law
\cite{Wigner:1948} only at very low collision energies (below
100 $\mu$K). At higher energies it actually decreases {\em
faster} than predicted by the threshold law and the elastic
cross section also decreases. The p-wave contribution to the
inelastic cross section is non-resonant, and increases
approximately as $E^{+1/2}$ with energy while the s-wave
contribution decreases. For the resonance in Fig.\
\ref{fig-reso2-1} its value is about 1~\AA$^2$ at 10~mK, so
that p-wave scattering will dampen the suppression of inelastic
scattering at temperatures above this.

The resonances for He + O$_2$ are quite wide, with
$|\Gamma_B|=250$ to 500 G. The resonance in Fig.\
\ref{fig-reso2-1} provides substantial suppression of inelastic
cross sections across a range of at least 100~G. This would be
sufficient to provide a working energy range of about 10~mK for
magnetically trapped states of a molecule in a $^3\Sigma$
electronic state. The broad resonances observed here contrast
with those previously characterized for He-NH, which had
$|\Gamma_B|<10^{-2}$ G \cite{Gonzalez-Martinez:2007}. The
difference arises in this case because the $n=1$ closed
channels involved for O$_2$ are directly coupled to the
inelastic channel(s) by the potential anisotropy, whereas the
$n=0$ closed channels involved for NH were only indirectly
coupled to the open channels. However, broad resonances are
likely to be common in molecule-molecule systems, and indeed in
atom-molecule systems involving atoms heavier than He, because
the couplings due to potential anisotropy are much stronger in
heavier systems \cite{Zuchowski:NH3:2008, Soldan:MgNH:2009}.

The asymmetric lineshapes observed here are analogous to Fano
lineshapes \cite{Fano:1961} in bound-free absorption spectra.
Fano considered the interference between the bound and
continuum contributions to a transition matrix element near
resonance. He showed that the bound-state contribution rises
from zero to a peak at resonance while the continuum
contribution drops from its background value to zero and
changes sign at resonance. When there is only a single
continuum channel, there is always a point near resonance where
the bound and continuum contributions cancel completely.
However, when there are $N$ outgoing channels, there is one
particular linear combination of them that is coupled to the
bound state and $N-1$ orthogonal linear combinations that are
not \cite{Fano:1961}. The resonance suppresses inelastic
scattering into the former but not into the latter, so the
cross section does not drop to zero.

For low-energy resonant scattering in the presence of inelastic
channels, the partial width for the incoming (elastic) channel
is proportional to the incoming wavevector $k_0$, while the
partial widths for the inelastic channels are essentially
independent of $k_0$ \cite{Hutson:res:2007}. At low energies we
may therefore consider the bound state to be coupled only to
the outgoing (inelastic) channels and apply Fano theory
directly to the inelastic cross sections.

Even for $^{16}$O$_2$ molecules at the $j=2,m_j=-2$ threshold,
which can relax only to form $j=0,m_j=0$, there are outgoing
channels with several values of $L'$. For s-wave scattering
($L=0$), $L'$ must be at least $\Delta m_j$ and must be even to
conserve parity. The kinetic energy release of 1.9 K at
$B=9750$~G is above the centrifugal barrier for $L=2$ (0.4 K)
but below that for $L=4$ (2.4 K). Because of this, the $L=2$
channel dominates the inelastic scattering away from resonance
{\em and} is the outgoing channel most strongly coupled to the
bound state. The inelastic cross section therefore shows a deep
minimum, though there is still a little background inelastic
scattering that is not suppressed by the resonance.

The situation is somewhat different for resonances at the
$j=2,m_j=-1$ threshold, such as that shown in Fig.\
\ref{fig-reso2-2} (resonance 2 in Fig.\ \ref{fig-levs-B}). In
this case the resonance suppresses the total inelastic cross
section by less than a factor of 10 from its background value.
At 11660 G the kinetic energy release is 0.55 K for relaxation
to the $j=2,m_j=-2$ (upper) threshold and 2.3 K for relaxation
to the $j=0,m_j=0$ (lower) threshold. The $L=2$ outgoing
channels at both inelastic thresholds contribute significantly
to the inelastic scattering far from resonance. However, the
resonant bound state (with $m_j=+2$) is coupled much more
strongly to $j=0,m_j=0$ channels than to $j=2,m_j=-2$ channels.
The resonance therefore suppresses inelastic scattering into
the lower channel but there is significant background inelastic
scattering into the upper channel that is unaffected by the
resonance.

The kinetic energy release needed to surmount centrifugal
barriers depends on the reduced mass and will be smaller in
heavier systems than for He + O$_2$, as will the temperature
range in which s-wave scattering dominates. However, much lower
temperatures are already being achieved in current experiments
\cite{Ni:KRb:2008, Danzl:ground:2009}.


We conclude that inelastic cross sections may sometimes be
reduced dramatically by tuning near to a Feshbach resonance.
This may be very important for attempts to produce ultracold
molecules by evaporative or sympathetic cooling: applying a
suitable bias field could suppress inelastic collisions near
the bottom of a trap and allow cooling in cases where it would
otherwise be prevented by inelastic losses. The reduction may
occur for any atom or molecule in an internally excited state,
but is most dramatic when there is a single outgoing channel
that dominates the inelastic scattering and is strongly coupled
to the resonant channel. A common example of this will be
systems in which all but one of the outgoing channels are
suppressed by centrifugal barriers.


The authors are grateful to the Royal Society for an
International Joint Project grant and to EPSRC for funding the
collaborative project CoPoMol under the ESF EUROCORES programme
EuroQUAM.

\bibliography{../../all}

\end{document}